\documentclass[aps,prd,a4paper,twocolumn,showpacs,nofootinbib]{revtex4}
\usepackage{amssymb,amsmath,latexsym,mathrsfs}
\usepackage{graphicx,subfigure}
\usepackage{epsfig}
\usepackage{varioref,xr-hyper}
\usepackage{color}

\usepackage[hypertex,draft=false,verbose=false,debug=false
            hyperindex=true,hypertexnames=true]{hyperref}
\begin{document}

\title{New Constraints on variations of the fine structure constant from CMB anisotropies}

\author{Eloisa Menegoni$^{a}$}
\author{Silvia Galli$^{a,b}$}
\author{James G. Bartlett$^{b,c}$}
\author{C.J.A.P. Martins$^{d,e}$}
\author{Alessandro Melchiorri$^{a}$}

\affiliation{$^a$ Physics Department and INFN, Universit\`a di Roma ``La Sapienza'', Ple Aldo Moro 2, 00185, Rome, Italy}
\affiliation{$^b$ Laboratoire Astroparticule et Cosmologie (APC), Universit\'e Paris Diderot, 75205 Paris cedex 13}
\affiliation{$^c$ Jet Propulsion Laboratory, California Institute of Technology, California, USA}
\affiliation{$^d$ Centro de Astrof\'{\i}sica, Universidade do Porto, Rua das Estrelas, 4150-762 Porto, Portugal}
\affiliation{$^e$ DAMTP, University of Cambridge, Wilberforce Road, Cambridge CB3 0WA, United Kingdom}

\begin{abstract}
We demonstrate that recent measurements of Cosmic Microwave Background temperature and polarization
anisotropy made by the ACBAR, QUAD and BICEP experiments substantially improve the cosmological constraints on
possible variations of the fine structure constant in the early universe. This data, combined with the five year
observations from the WMAP mission yield the constraint $\alpha / \alpha_0 = 0.987 \pm 0.012$ at $68 \%$ c.l..
The inclusion of the new HST constraints on the Hubble constant further increases the accuracy to $\alpha / \alpha_0 = 1.001 \pm 0.007$
at $68 \%$ c.l., bringing possible deviations from the current value below the $1 \%$ level and improving previous
constraints by a factor $\sim 3$.
\end{abstract}

\maketitle

\section{Introduction}

Nature is characterized by a number of physical laws and fundamental dimensionless couplings, which historically we have assumed to
be spacetime-invariant. For the former this assumption is a cornerstone of the scientific method (and alternatives are virtually
inconceivable), but for the latter it is an assumption with no real justification. There is ample experimental evidence showing that
fundamental couplings run with energy, and many particle physics and cosmology models suggest that they should also roll with time.

Searching for time variations of fundamental constants (see e.g. \cite{uzan} and references therein) is a challenging but powerful probe of
fundamental physics, and starts with the identification of laboratory or astrophysical environments that are so 'clean' and well understood that
any deviations from the expected behavior can be ascribed to new physics (as opposed to systematics or other uncertainties). The Cosmic
Microwave Background (CMB, hereafter) is one such example.

The astonishing agreement between the current measurements of and the theoretical expectations for CMB temperature and polarization anisotropies has
opened the possibility of testing several aspects of fundamental physics in the early universe (see e.g. \cite{wmap5cosm,wmap5komatsu}). In particular,
CMB anisotropies are sensitive to variations in fundamental constants such as the fine structure constant $\alpha$
(see e.g. \cite{hannestad}, \cite{kaplinghat}, \cite{battye}, \cite{avelino}).

A time varying fine structure constant can leave an imprint on CMB anisotropies by changing the time of recombination and the size of the
acoustic horizon at photon-electron decoupling, and the steadily improving CMB datasets have been extensively used to constrain it.
Parameterizing a variation in the fine structure constant as $\Delta_\alpha=(\alpha-\alpha0)/\alpha_0$, where
$\alpha_0=1/137.03599907$ is the standard, local, value and $\alpha$ is the value during the recombination process, the authors of \cite{rocha} used
the first year WMAP data, finding the constraint $-0.06<\Delta_\alpha<0.01$ at $95 \%$ c.l. (see also \cite{ichikawa}).
This constraint was subsequently updated to $-0.039<\Delta_\alpha<0.01$ (see \cite{petruta}) by combining the third year
WMAP data with the Hubble Space Telescope key project constraint on the Hubble Constant.
More recently, using the five observations from the WMAP satellite, the authors of \cite{jap} found the constraint $-0.05 < \Delta_\alpha < 0.042$.
It is well known (see e.g. \cite{petruta}) that a variation in the fine-structure constant is mostly
degenerate with a variation in the Hubble constant $H_0=100hKm/s/Mpc$. Combining CMB data with
independent measurements of $H_0$ can indeed improve the constraint on $\alpha$.

Over the past few months substantial improvements have been reported both in measurements of CMB anisotropies and in the
determination of the Hubble constant. The new results from the ACBAR (\cite{acbar}), QUAD (\cite{quad}) and BICEP (\cite{bicep}) experiments,
together with the  new release of the WMAP data from the five-year survey, are now sampling the CMB temperature angular spectrum with
great accuracy down to arcminute angular scales and now also provide clear evidence for acoustic oscillations in the polarization channel.
Moreover, the uncertainty on the Hubble constant has been reduced by more than half from the
recent analysis of \cite{riess} yielding a new constraint of $h=0.742 \pm 0.036$.

With these new experimental improvements it is therefore timely to investigate the new constraints on
$\alpha$, as we plan to do in this brief report. Our paper is therefore structured as follows:
in the next Section we briefly describe the data analysis method used while in Section III we present our results.
As we show in the Conclusion, the new data provides a significant improvement on the constraint on a time varying $\alpha$.

\section{Analysis Method}
\label{3}

We include a possible variation in the fine structure constant in the
recombination process using the method adopted in \cite{avelino} and
modifying the publicly available RECFAST (\cite{recfast}) routine
in the CAMB (\cite{camb}) CMB code.

We constrain variation in the fine structure constant $\alpha / \alpha_0$
by a COSMOMC analysis of the most recent CMB data. The analysis method we adopt is based on the
publicly available Markov Chain Monte Carlo package \texttt{cosmomc}
\cite{Lewis:2002ah} with a convergence diagnostics done through the Gelman and Rubin statistics.

We sample the following eigh-dimensional set of cosmological parameters, adopting flat priors
on them: the baryon and cold dark matter densities $\omega_{\rm b}$ and
$\omega_{\rm c}$, the Hubble constant $H_0$, the scalar spectral index $n_s$,
the overall normalization of the spectrum $A_s$ at $k=0.05$ Mpc$^{-1}$,
the optical depth to reionization, $\tau$ and, finally, the variations in the
fine structure constant $\alpha / \alpha_0$.
Furthermore, we consider purely adiabatic initial conditions and we impose spatial flatness.

Our basic data set is the five--year WMAP data \cite{wmap5cosm}, \cite{wmap5komatsu}
(temperature and polarization) with the routine for computing the likelihood supplied by the WMAP team.
Together with the WMAP data we also consider the following CMB datasets:
ACBAR (\cite{acbar}), QUAD (\cite{quad}) and BICEP (\cite{bicep}).
We also include the older datasets from BOOMERanG (\cite{boom03}) and CBI (\cite{cbi}).
For all these experiments we marginalize over a possible contamination
from Sunyaev-Zeldovich component, rescaling the WMAP template at the corresponding
experimental frequencies. Finally, we include the improved constraint on the Hubble constant of $h=0.747\pm0.036$
at $68 \%$ c.l.. from the recent analysis of \cite{riess}.

\section{Constraints on variations of the fine structure constant}
\label{4}

In Table \ref{tab:wdm} we report the constraints on the $\alpha /\alpha_0$ parameter obtained from the COSMOMC analysis, using the
the different combinations of the datasets described in the previous section.

\begin{table}[h!]
\begin{center}
\begin{tabular}{lclll}
Experiment & & $\alpha / \alpha_0$ & $68\%$ c.l. & $95\%$ c.l. \\
\hline
\vspace{0.2cm}
WMAP-$5$ & & $0.998$ & $\pm 0.021$ & ${}^{+0.040}_{-0.041}$\\
\vspace{0.2cm}
All CMB & & $0.987$ & $\pm 0.012$ & $\pm 0.023$\\
\vspace{0.2cm}
All CMB+ HST & & $1.001$ & $\pm 0.007$ & $\pm 0.014$\\
\hline
\end{tabular}
\caption{Limits on $\alpha / \alpha_0$ from WMAP data only (first row), from a larger set of CMB experiments (second row),
and from CMB plus the HST prior on the Hubble constant, $h=0.748\pm0.036$ (third row).
We report errors at $68\%$ and $95\%$ confidence level.}
\label{tab:wdm}
\end{center}
\end{table}

Clearly the new CMB data at arcminute angular scales provide a substantial improvement in the determination of
$\alpha$. The uncertainty on $\alpha$ is indeed halved when the data coming from the new QUAD, BICEP and ACBAR experiments
are included in the analysis. The increase in the precision is mainly due to the effects from modified recombination on
the CMB anisotropy damping tail that is more accurately measured by the QUAD and ACBAR experiments.

\begin{figure}[h!]
\includegraphics[width=7.2cm]{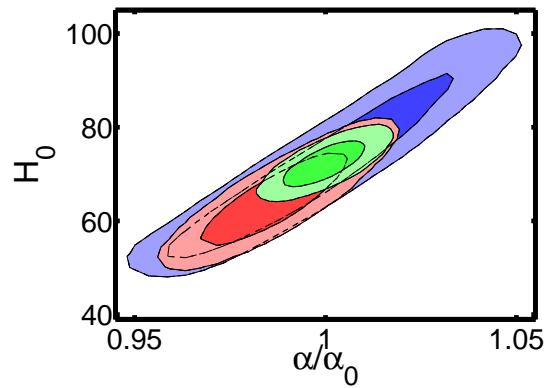}
\caption{\label{plot}$68 \%$ and $95 \%$ c.l. constraints on the $\alpha / \alpha_0$ vs
the Hubble constant for different datasets. The contour regions come from the WMAP-$5$ data (blue),
all current CMB data (red), and CMB+HST (green).}
\end{figure}

In Figure \ref{plot} we show the $68 \%$ and $95 \%$ c.l. constraints on the $\alpha / \alpha_0$ vs
the Hubble constant for different datasets. As we can see, a degeneracy is clearly present between the Hubble
parameter and the fine structure constant. A change in $\alpha$ shifts the recombination epoch, affecting the
angular diameter distance at recombination and the peak position in the CMB anisotropy angular spectra.
A similar effect can be obtained by changing the value of the Hubble constant and the two parameters are
therefore degenerate. Including the recent HST measurements of $H_0$ has therefore the important effect
of breaking the $\alpha$-$H_0$ degeneracy and thereby providing a stronger bound to $\alpha /\alpha_0$. This is
clearly shown in the third row of Table 1 and, again, in Figure \ref{plot}.
The best fit parameters for the CMB+HST run are: $\Omega_bh^2=0.0228$, $\Omega_ch^2=0.112$, $\tau=0.093$,
$h=0.720$, and $n_s=0.964$, in agreement with the values obtained when $\alpha$ is not varied.

\begin{figure}[h!]
\includegraphics[width=7.2cm]{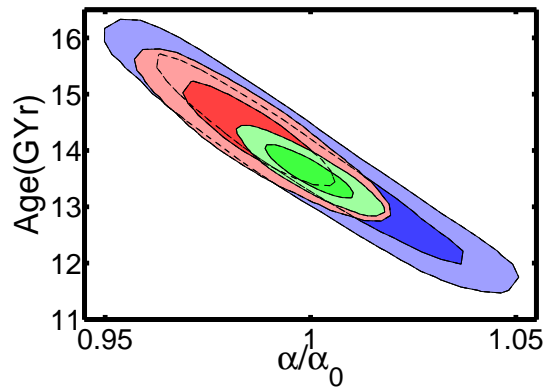}
\caption{\label{plot2}$68 \%$ and $95 \%$ c.l. constraints on the $\alpha / \alpha_0$ vs
the age of the Universe $t_0$ for different datasets. The contour regions come from the WMAP-$5$ data (blue),
all current CMB data (red), and CMB+HST (green).}
\end{figure}

Since the age of the universe is strongly connected with the Hubble constant, we also plot the constraints
on the $\alpha / \alpha_0$ vs age of the universe plane in Figure \ref{plot2}. As we
can see, including the small scale CMB experiments and the HST prior recover the standard constraint on the
Age of the universe achieved assuming constant $\alpha$. We indeed found that if one allows for variations in
$\alpha$, the WMAP five year data bounds the age of the universe to $t_0=13.9\pm1.1$ Gyrs (at $68 \%$ c.l.),
with an increase in the error of a factor $\sim 3$ respect to the quoted standard constraint (see \cite{wmap5cosm}).
Including all CMB datasets improves the constraint to $t_0=14.3\pm0.6$ while combining with the HST prior yields
$t_0=13.6\pm0.3$ Gyrs (all at $68 \%$ c.l..).

We found no relevant degeneracies with the remaining parameters.

\section{Conclusions}

In this brief report we have investigated the constraints on variation of the fine structure constant from
current CMB observations. We have updated previous results and investigated in detail the degeneracies
present between $\alpha$ and the remaining cosmological parameters.

We have found that the combination of the latest CMB data and HST measurement of $H_0$  yield the
constraint $\delta \alpha / \alpha = 1.000 \pm 0.007$ at $68 \%$ c.l., providing no indications for
strong variation in $\alpha$ in the early universe. This constraint improves by a factor
$\sim 3$ previous bounds obtained by past cosmological data analysis. This bound is also
competitive with Big Bang Nucleosynthesis bounds (see e.g. \cite{avelino}) which assume the same cosmological framework,
but is based on different physical mechanisms at very different energy scales. (Stronger BBN bounds can be obtained by making specific
model-dependent assuptions \cite{Iocco}.)

Further improvements are expected from the Planck satellite experiment, which is now collecting data \cite{rocha}: it should be able to bound
variations in $\alpha$ at the $\sim 0.5 \%$ level without assuming the HST prior (for comparison a cosmic variance limited experiment could
improve the bound to $\sim 0.1 \%$ level). These ever tighter bounds, combined with local measurements using atomic clocks and with forthcoming
low-redshift measurements obtained with stable high-resolution spectrographs such as PEPSI, ESPRESSO or CODEX, strongly constrain fundamental physics,
particularly the dynamics of any cosmological scalar fields.

\begin{acknowledgments}
The work of C.M. is funded by a Ci\^encia2007 Research Contract, supported by FSE and POPH-QREN funds.
\end{acknowledgments}


\begin{thebibliography}{99}

\bibitem{uzan}
  J.~P.~Uzan,
  Rev.\ Mod.\ Phys.\  {\bf 75} (2003) 403
  [arXiv:hep-ph/0205340];
  C.~G.~Scoccola,
  arXiv:0906.0329 [astro-ph.CO].

\bibitem{wmap5cosm}
  G.~Hinshaw {\it et al.}  [WMAP Collaboration],
  arXiv:0803.0732 [astro-ph].

\bibitem{wmap5komatsu}
  E.~Komatsu {\it et al.},
  arXiv:0803.0547 [astro-ph].

\bibitem{hannestad}
  S.~Hannestad,
  Phys.\ Rev.\  D {\bf 60} (1999) 023515
  [arXiv:astro-ph/9810102].

\bibitem{kaplinghat}
  M.~Kaplinghat, R.~J.~Scherrer and M.~S.~Turner,
  Phys.\ Rev.\  D {\bf 60} (1999) 023516
  [arXiv:astro-ph/9810133].

\bibitem{battye}
  R.~A.~Battye, R.~Crittenden and J.~Weller,
  Phys.\ Rev.\  D {\bf 63} (2001) 043505
  [arXiv:astro-ph/0008265].

\bibitem{avelino}
  P.~P.~Avelino {\it et al.},
  Phys.\ Rev.\  D {\bf 64} (2001) 103505
  [arXiv:astro-ph/0102144];
 C.~J.~A.~Martins, A.~Melchiorri, R.~Trotta, R.~Bean, G.~Rocha, P.~P.~Avelino and P.~T.~P.~Viana,
  Phys.\ Rev.\  D {\bf 66} (2002) 023505
  [arXiv:astro-ph/0203149].

\bibitem{rocha}
  C.~J.~A.~Martins, A.~Melchiorri, G.~Rocha, R.~Trotta, P.~P.~Avelino and P.~T.~P.~Viana,
  Phys.\ Lett.\  B {\bf 585}, 29 (2004)
  [arXiv:astro-ph/0302295];
  G.~Rocha, R.~Trotta, C.~J.~A.~Martins, A.~Melchiorri, P.~P.~Avelino, R.~Bean and P.~T.~P.~Viana,
  Mon.\ Not.\ Roy.\ Astron.\ Soc.\  {\bf 352}, 20 (2004)
  [arXiv:astro-ph/0309211].

\bibitem{ichikawa}
  K.~Ichikawa, T.~Kanzaki and M.~Kawasaki,
  Phys.\ Rev.\  D {\bf 74} (2006) 023515
  [arXiv:astro-ph/0602577].

\bibitem{petruta}
  P.~Stefanescu,
  New Astron.\  {\bf 12} (2007) 635
  [arXiv:0707.0190 [astro-ph]].

\bibitem{jap}
  M.~Nakashima, R.~Nagata and J.~Yokoyama,
  Prog.\ Theor.\ Phys.\  {\bf 120} (2008) 1207
  [arXiv:0810.1098 [astro-ph]].

\bibitem{acbar}
  C.~L.~Reichardt {\it et al.},
  Astrophys.\ J.\  {\bf 694} (2009) 1200
  [arXiv:0801.1491 [astro-ph]].

\bibitem{quad}
  M.~L.~Brown {\it et al.}  [QUaD collaboration],
  arXiv:0906.1003 [astro-ph.CO].

\bibitem{bicep}
  H.~C.~Chiang {\it et al.},
  arXiv:0906.1181 [astro-ph.CO].

\bibitem{riess}
  A.~G.~Riess {\it et al.},
  Astrophys.\ J.\ Suppl.\  {\bf 183} (2009) 109
  [arXiv:0905.0697 [astro-ph.CO]].

\bibitem{recfast}
  W.~Y.~Wong, A.~Moss and D.~Scott,
  arXiv:0711.1357 [astro-ph].

\bibitem{camb}
  A.~Lewis, A.~Challinor and A.~Lasenby,
  Astrophys.\ J.\  {\bf 538} (2000) 473
  [arXiv:astro-ph/9911177].

\bibitem{Lewis:2002ah}
A. Lewis and S. Bridle,
Phys.\ Rev.\ D {\bf 66}, 103511 (2002) (Available from
\texttt{http://cosmologist.info}.)

\bibitem{boom03}
  W.~C.~Jones {\it et al.},
  arXiv:astro-ph/0507494;  F.~Piacentini {\it et al.},
  arXiv:astro-ph/0507507;
  arXiv:astro-ph/0507514.

\bibitem{cbi}
A.~C.~S.\ Readhead {\em et al.},
Astrophys.\ J.\  {\bf 609}, 498 (2004).

\bibitem{Iocco}
  F.~Iocco, G.~Mangano, G.~Miele, O.~Pisanti and P.~D.~Serpico,
  Phys.\ Rept.\  {\bf 472} (2009) 1
  [arXiv:0809.0631 [astro-ph]].


\end{thebibliography}
\end{document}